\documentclass{aa}  
\usepackage{graphicx}
\usepackage{natbib}
\bibpunct{(}{)}{;}{a}{}{,} % to follow the A&A style  
\newcommand{\isaac}{{\sl ISAAC}}
\newcommand{\vlt}{{\sl VLT}}
\newcommand{\cxo}{{\sl Chandra}}
\newcommand{\xmm}{{\sl XMM}}
\newcommand{\twomass}{{\sl 2MASS}}
\newcommand{\vltn}{{\sl Very Large Telescope}}
\begin{document}
   \title{Deep VLT infrared observations of X-ray Dim Isolated Neutron Stars\thanks{Based on archive data from
 observations collected at the European Southern Observatory, Paranal, Chile under programs ID 071.C-0189(A), 
 072.C-0051(A) and 074.C-0596(A) }}

\titlerunning{IR observations of XDINSs}

%   \subtitle{}

   \author{G.~Lo~Curto \inst{1}
	  \and
          R.P.~Mignani \inst{2}
	  \and
	  R.~Perna \inst{3}
	  \and
	  G.L.~Israel \inst{4}
          }

   \offprints{G.~Lo~Curto,\\ e-mail: {\tt glocurto@eso.org} }

   \institute{European Southern Observatory, Av. Alonso de Cordova 3107, Santiago, Chile
	\and
	 Mullard Space Science Laboratory, University College London, Holmbury St. Mary, Dorking, Surrey, RH5 6NT, UK 
	\and
	JILA and Department of Astrophysical and Planetary Sciences, University of Colorado, 440 UCB, Boulder, 80309, USA
	\and
	INAF Astronomical Observatory of Rome, Via Frascati 33, 00040, Monte Porzio Catone, Italy 
	}

   \date{Received ; accepted }

% \abstract{}{}{}{}{} 
% 5 {} token are mandatory
 
  \abstract
   {
X-ray observations have unveiled the existence of a family of
radio-quiet Isolated Neutron Stars whose X-ray emission is purely 
thermal, hence dubbed X-ray Dim Isolated Neutron Stars (XDINSs).
While  optical
observations  have  allowed to  relate  the  thermal  emission to  the
neutron star  cooling and  to build the  neutron star  surface thermal
map,  IR observations  are critical  to pinpoint  a  spectral turnover
produced  by a  so  far  unseen magnetospheric  component,  or by  the
presence of  a fallback  disk. 
The detection of such a turnover can provide further evidence of a
link  between  this  class of  isolated
neutron  stars and the  magnetars, which  show a  distinctive spectral
flattening in the IR.
}
   {
    We present the deepest IR observations ever of five XDINSs,
    which we use to constrain  a spectral turnover
    in the IR and the presence of a fallback disk.
   }
   {
We have used archived VLT observations of these neutron stars
performed with  the ISAAC instrument  in the $H$-band (1.65  $\mu$m) and
the available fallback disk models.
   }
   {
For  none  of  our targets it  was  possible  to  identify  the  IR
counterpart  down to limiting  magnitudes $H\sim  21.5-22.9$. Although
these limits are  the deepest ever obtained for  neutron stars of this
class, they  are not deep  enough to rule  out the existence  and the
nature of  a possible spectral  flattening in the  IR. 
We also derive, by using disk models, the upper limits on the mass 
inflow rate in a fallback disk.
We find the existence of a putative fallback disk consistent 
(although not confirmed) with our observations.
   }
{}
%   {We conclude that,  our observations are consistent with the age and disk mass estimates}

   \keywords{Isolated Neutron Stars -- XDINS --
                Pulsars --
                Near Infrared
               }

   \maketitle
%
%________________________________________________________________

\begin{table*}
\centering                            % used for centering table
\begin{tabular}{c c c c c l}        % centered columns (4 columns)
\hline\hline                          % inserts double horizontal lines
Target & Date & Exp. time & seeing & airmass & sky  \\
       & dd-mm-yyyy & (s) & (arcsec) & &\\
\hline                                % inserts single horizontal line
RX J0420.0--5022 & 14-11-2003 & 1980 & 0.8 & 1.4 & clear \\
		& 11-01-2004 & 1980 & 0.6 & 1.3 & clouds \\
		& 14-01-2004 & 1980 & 1.1 & 1.7 & cyrrus \\

RX J0720.4--3125 & 05-12-2003 & 1980 & 0.8 & 1.0 & photometric  \\
		& 09-01-2004 & 1980 & 0.6 & 1.1 & cyrrus  \\
		& 12-01-2004 & 1980 & 0.7 & 1.0 & clear \\

RX J0806.4--4123 & 26-11-2004 & 3960 & 0.8 & 1.0 & clear   \\
		& 22-12-2004 & 1980 & 0.6 & 1.1 & clear   \\

RX J1856.5--3754 & 23-05-2003 & 2760 & 0.6 & 1.1 & cyrrus \\
RX J2143.0+0654 & 07-06-2003 & 2580 & 0.5 & 1.2 & cyrrus\\
\hline                                 %inserts single line
\end{tabular}
\caption{Summary of the \isaac\ observations of our targets. Column two gives the observing dates(s), columns three to five give the total integration time per observation, the average seeing and airmass. Column six gives the sky conditions.    }
\label{conditions}      % is used to refer this table in the text
\end{table*}

\section{Introduction}
Observations  performed  at $\gamma$-ray  and  X-ray energies  have
unveiled the  existence of peculiar classes of  Isolated Neutron Stars
(INSs)  which stand  apart from  the  family of  more classical  radio
pulsars as  they are, with only one confirmed exception \citep{camilo2006}, radio-quiet.  The 
Soft Gamma-ray  Repeaters (SGRs) and the Anomalous X-ray Pulsars 
(AXPs) are believed to be magnetars, i.e.  young
neutron stars  with long spin periods (4--11 s) and hyper-strong magnetic
fields \citep[$B \sim 10^{14}-10^{15}$ G, see][]{woods2006}.  
The magnetar model \citep{duncan1992} would explain
several observational properties of  SGRs and AXPs, 
including the period, the spin down rate and their luminosity,
in excess to what is expected from a standard pulsar spin down,
see \citet{woods2006} for a recent review. 
%including the lack
%of radio  emission, which  may be quenched  by magnetic  fields higher
%than the  quantum critical  value $B_c =  4.41 \times 10^{13}$  G, 
%on  AXPs and SGRs.  
Other radio-quiet  INSs  are  the  dim  X-ray  sources  detected  by  ROSAT,
hereafter X-ray  Dim INSs  or XDINSs, see \citet{haberl2006} for  a review. \\
The XDINSs are characterized by steady emission patterns in the soft X-rays 
regime, where the spectra can be described by black-bodies with
$40\leq KT \leq 100$ eV and by high X-ray to optical flux ratio 
($log(f_{{\rm X}}/f_{{\rm opt}})\approx 4 - 6$). 
%
%One of the main goals in studying neutron stars is the understanding
%of the equation of state of matter at very high density. XDINSs are an 
%ideal test case as their thermal X-ray spectra come directly from their
%surface and their interpretation is not complicated by modeling the 
%magnetosphere of the neutron star.
The XDINSs soft X-ray emission is likely powered by the cooling of the 
neutron star and it is pulsed with periods in the  3--11 s range.

%According to their thermal spectra and cooling models \citep{page2006} 
According to cooling models \citep{page2006}, the inferred surface temperatures 
imply that XDINSs ages are in the range of $\sim 10^5 - 10^{6}$ years
which, in the case of RX J1856.6--3754 and RX J0720.4--3125, 
is compatible with their place of origin being in the Upper Scorpio 
and in the Trumpler 10 OB associations respectively, once their present
position and proper motion are taken into account \citep{kaplan2007}.

There  is evidence  which suggest  that magnetars  and XDINSs  may be
linked at  some level. They have  similar X-ray periods,  and for four
XDINSs,  RX J1308.6+2127 \citep{haberl2003},  RX J1605.3+3249 \citep{kerk2004},
RX J0720.4--3125 \citep{haberl2004} and RX J2143.0+0654 \citep{zane2005}  
%the  measured  cyclotron X-ray  absorption features 
the  observed X-ray absorption features, due either to proton cyclotron 
resonance or to atomic transitions,
yield magnetic fields $B  \sim 6-7 \times 10^{13}- 1.0 \times
10^{14}$G.   For  RX J0720.4--3125  and  RX J1308.6+2127 the  measured
period derivatives  $\dot P  \sim 7 \times  10^{-14}$ s  s$^{-1}$ and
$\dot P \sim  11 \times 10^{-14}$ s s$^{-1}$ 
\citep{kap2005a, kap2005b}  imply, assuming  magnetic dipole spin  down, 
$B  \sim 3 \times 10^{13}$ G, i.e. consistent with the values derived from X-ray spectroscopy.  In addition, for the measured parallactic distance 
of 360 pc \citep{kaplan2007} the X-ray luminosity of RX J0720.4--3125 turns
out to be larger than  its inferred spin down energy, a characteristic
which makes it similar to the magnetars which have an X-ray luminosity
by at least  2 orders of magnitudes higher with  respect to their spin
down energy.  In  the IR, magnetars have a  peculiar flux distribution
which flattens with respect to  the extrapolation of the X-ray spectra,
see e.g. \citet{israel2005}. This flattening 
could be  due either to  the contribution of emission from a fallback disk,
or to a turnover in the magnetospheric  emission from  the  magnetar.
In turn, the emission from the fallback disk could be generated by mass transfer 
and viscous dissipation within the disk and/or by re-processing of the 
X-rays from the neutron star.
Were the XDINSs show a similar energy distribution as the magnetars, flattening
their spectra in the IR, it would strengthen the link between these two 
classes of objects.
%As we do not measure magnetospheric emission from the XDINSs, finding 
%a similar  behavior would point to the presence of 
%an accreting fallback disk as the cause for this emission, both for 
%XDINSs and magnetars.

In this paper  we report on recent IR observations  of five XDINSs: RX
J0420.0--5022, RX J0720.4--3125, RX J0806.4--4123, RX J1856.6--3754 and RX
J2143.6+0654.  The paper is  divided as follows: observations, data
reduction and results are described  in \S  2,  while the discussion is
presented in \S 3.

%__________________________________________________________________

\section{Observations, Data Reduction and Results}
\label{obs}

%\subsection{Observations}
%\label{obs}

XDINSs  IR observations  were conducted  using the  NIR spectra-imager
\isaac\ \citep{ISAAC} mounted at the First Unit Telescope (UT1) of the
ESO \vltn\  (\vlt) at the  Paranal Observatory in Chile  under program
IDs 071.C-0189(A),  072.C-0051(A) and 074.C-0596(A), PIs R. Neuh\"auser 
and B. Posselt.  The instrument
configuration was  set to the  Short Wavelength (SW)  camera, equipped
with a  Rockwell Hawaii  1024$\times$1024 pixel Hg:Cd:Te  array, which
has a pixel  size of 0\farcs148 and a field  of view of 152$\times$152
arcsec. All  observations were  performed through the  $H$-band filter
($\lambda=  1.65  \mu m;  \Delta  \lambda=  0.30  \mu m$). To  allow  for
subtraction of the variable IR sky background, each exposure was split
in  sequences of  14  to  33 short  randomly  dithered exposures  with
Detector Integration Times (DIT) of 12 s, and 5 NDIT repetitions along
each  point  of the  dithering  pattern.  Observations were  performed
between May 2003 and December 2004. Table \ref{conditions} reports the
observations  summary.  Science exposures  were retrieved  through the
public ESO archive\footnote{http://archive.eso.org/} together with the
closest in time calibration files.

The data were reduced using the ESO's eclipse
package \citep{eclipse}\footnote{http://www.eso.org/projects/aot/eclipse/eug/eug/eug.html}
for  de-jitter  and sky  subtraction.   
%
%__________________________________________________________________
%
%\subsection{Photometry}
%\label{phot}
%
Unfortunately, we
could not  find in the ESO  archive IR standard stars  observed sufficiently 
close in date and time to our targets. For this reason, we have used the $H$-band
flux  of a  number  of \twomass\  stars  identified in  the frames  as
relative  calibration  sources.   Since  the photometric  accuracy  of
\twomass\  is as good  as $\sim  0.1$ magnitudes,  and since  a direct
on-the-frame photometric calibration is  not affected by overnight sky
background  fluctuations,  we  conclude   that  the  accuracy  of  our
photometric  calibration  is fully satisfactory. We have used all
available \twomass\ stars in each frame (between 3 and 7) for the relative 
photometric calibration.
%comparable  to  the  one  which would  be
%obtained using  twilight photometric standards.   
Also, the wavelength
overlap of  the \twomass\ and  the \isaac\ $H$-band filters  is almost
complete  and the  introduced error  is within  0.03  magnitudes.
% i.e., comparable with the photon noise affecting a single measurement. 
Photometry on the  \twomass\ stars was performed via standard
aperture photometry with the Skycat - GAIA package \citep{gaia}. Frame
distortions  (amounting up  to 2  pixels and depending on position of the 
source in the frame) were  not corrected  for, and introduce an  error of 
up to  0.1 magnitudes on the  photometry of the individual 
\twomass\  stars,   which  largely  dominates   over  the  statistical
errors. 
We  finally  derive  a  global  photometric  accuracy  of  $\sim  0.1 - 0.2$
magnitudes.

%__________________________________________________________________

%\subsection{Astrometry}
%\label{astro}

The astrometric  solution for the stacked \isaac\  images was computed
by  using  as a  reference  the  positions  and coordinates  of  stars
selected  from  the  \twomass\  catalogue,  yielding  an  rms  between
$0\farcs09$  and  $0\farcs14$.   To  register the  positions  of  our
targets  we have  used  as   reference  the  most precise  coordinates
available  in  the literature  (see  Table  \ref{positions}).  
For  RX J0420.0--5022 and RX  J0806.4--4123 we have  used the
X-ray coordinates obtained with \cxo\ \citep{haberl2004}, for RX J0720.4--3125 and RX
J1856.6--3754 we have used  the coordinates of the optical counterparts
after  correcting for  the measured  proper motions  
\citep[\citeauthor{mvk2006}, \citeyear{mvk2006} and][respectively]{vk2001}
while for RX J2143.0+0654 we use the coordinates obtained with \xmm\  \citep{zane2005}.
The overall uncertainties on the targets position
%registered coordinates  of our targets are shown in
%the  five panels  of Fig.   1,  where the  size of  the error  circles
account   for  the   errors  on   the  sky   coordinates   (see  Table
\ref{positions}),  for the  rms of  the astrometric  solution,  for the
absolute  uncertainty of the  \twomass\ coordinates  ($\sim 0\farcs2$)
and, for  RX J0720.4--3125 and RX  J1856.6--3754, for the  errors on the
proper motion  propagation between  the reference and  the observation
epochs.

\begin{table}
\centering
\begin{tabular}{c l l c l}        % centered columns (4 columns)
\hline\hline                 % inserts double horizontal lines
Target & R.A. & Dec. & $\Delta \: r$ & ref. \\
\hline                                   %inserts single line
RX J0420.0$-$5022 & 04 20 01.95 & $-$50 22 48.1 & 0\farcs6 & 1 \\
RX J0720.4$-$3125 & 07 20 24.96 & $-$31 25 50.1 & 0\farcs2 & 2 \\
RX J0806.4$-$4123 & 08 06 23.40 & $-$41 22 30.9 & 0\farcs6 & 1 \\
RX J1856.5$-$3754 & 18 56 35.62 & $-$37 54 35.3 & 0\farcs2 & 3 \\
RX J2143.0$+$0654 & 21 43 03.38 &$+$ 06 54 17.5 & 0\farcs6 & 4 \\

\hline                                   %inserts single line
\end{tabular}
\caption{Coordinates and position uncertainties of the observed objects. 
The uncertainties refer to a $90\%$ confidence level for the sources
not detected in optical (RX J0420.0--5022, RX J0806.4--4123, RX J2143.0+0654).
The reference epochs for the positions are detailed in the referenced papers:
1: \citet{haberl2004}, 2: \citet{mvk2006}, 3: \citet{vk2001}, 4: \citet{rea2007}.}
\label{positions}      % is used to refer this table in the text
\end{table}

   \begin{figure}
   \centering
\includegraphics[width=8cm,angle=0]{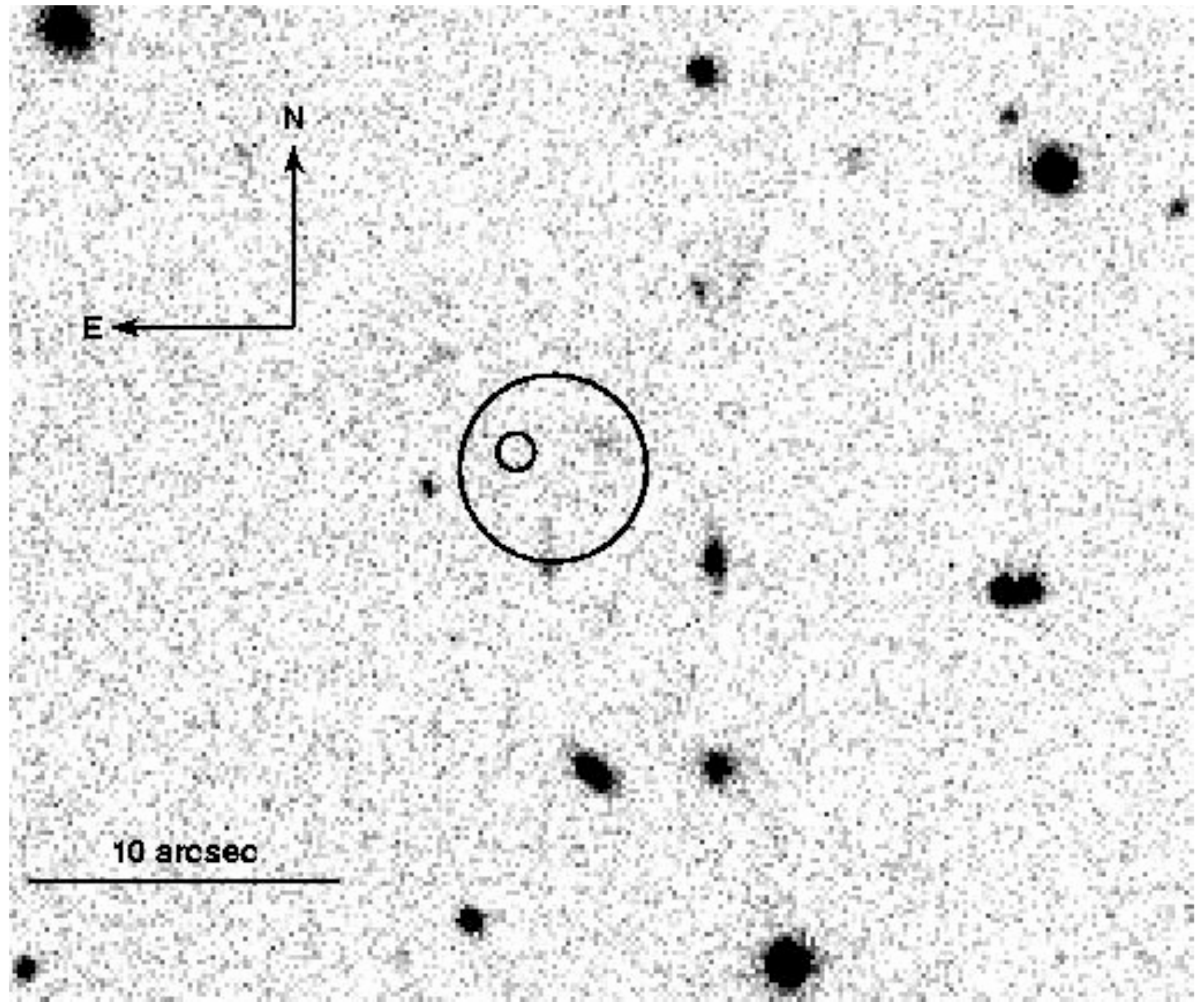}
	\caption{ISAAC  $H$-band  images of the field of RX J2143.6+0654.
		Although two sources appear within the large \xmm\ error circle, 
		no source is visible within the smaller \cxo\ error circle.
		Moreover the two sources within the \xmm\ circle have large values of B-V, 
		not consistent with the
		nature of neutron stars emission, and have optical luminosity 
		a factor 10 higher than expected \citep[see][and the text]{mig2006}.}
         \label{rxj2143}
   \end{figure}

%__________________________________________________________________

%\subsection{Results}
%\label{res}

We do not detect any candidate IR counterpart for four of our sources, even extending the 
search area to 3 error circle  radii (confirmed independently in \citeauthor{posselt2007} 
\citeyear{posselt2007}).
In the case of RX J2143.0+0654,
four possible  counterparts were already identified  in \citet{mig2006}
based on their  proximity with the \xmm\ large error circle (Fig. 1), but they were all
considered  unlikely candidates due  to their large $B-V > 0.5$,  to their
optical brightness, a factor 10 higher than expected and, 
at least  for two objects, to  their apparent
extension; the only point-like object within the \xmm\ error circle in the IR frame has a low
detection significance ($<5\sigma$).
Very recently the \cxo\ position was published for this object \citep{rea2007}.
No sources are visible within the smaller \cxo\ error circle.

%We therefore claim no detection for all of the sources.
The derived upper limits, at the $3 \sigma$ level, are shown in table 
\ref{results}.

\begin{table}
\begin{center}

\begin{tabular}{l|c|lcc} \hline
NS Name             & $H$               &d      & $N_H$   	      & $A_H$ \\ 
RX		    & magnitude	     & pc	& $10^{20}$ cm$^{-2}$ & mag. \\	
\hline
J0420.0--5022    &$> 21.7$  &$\approx 345$        &1.6	&0.016\\
J0720.4--3125    &$> 22.7$  &360$^{+170}_{-90}$   &1.2	&0.012\\
J0806.4--4123    &$> 22.9$  &$\approx 250$        &1.0	&0.010\\
J1856.5--3754    &$> 21.5$  &161$^{+18}_{-14}$	  &0.7	&0.007\\
J2143.0+0654     &$> 22.1$  &$\approx 430$        &2.4	&0.024\\ 
\hline
\end{tabular}
\caption
{
    Summary of the $H$-band flux upper limits (expressed in magnitude) for the five XDINSs.
    Limiting magnitudes are computed for a $3 \sigma$ detection level.
    Columns three and four give the neutron star distance and the $N_H$, column five gives the 
    interstellar extinction in the $H$-band, $A_H$, computed from the $N_H$ derived from the X-ray spectral 
    fits using  the relation of \citet{paresce84} as done by, e.g. \citet{posselt2006}. 
    We refer to \citet{posselt2006} for the estimation of $N_H$, and the distances
    of RX J0420.0--502, RX J0806.4--4123 and RX J2143.0+0654.
    For the objects with measured parallaxes, RX J0720.4--3125 and RX J1856.5--3754, 
    the distance measurements are taken from \citet{kaplan2007} and from \citet{mvk2006}, respectively.
}
\label{results}
\end{center}
\end{table}

%We thus reject these four objects as possible counterparts of RX J2143.0+0654.
%The updated  \cxo\  position definetely  rules out  their
%association with RX J2143.0+0654 (see  also Rea et al. 2007).  We thus
%conclude that our sources are undetected in the IR.

%__________________________________________________________________

\section{Discussion}

We note that our  derived IR  flux upper  limits are  well  above the
extrapolation of  the XDINSs' black body X-ray  spectra, 
e.g. \citet{haberl2006},  which   predicts  values  of  $H > 25$.
For RX J0720--3125 and RX J1856--3754,  the only two XDINSs for which a
reliable characterization  of the  optical flux distribution  has been
obtained \citep[\citeauthor{motch2003} \citeyear{motch2003}]{vk2001},  an
hypothetical spectral  flattening redward of the  $R$-band would imply
$H  \le 24$  and $H\le  22$,  respectively. 
The estimation is made assuming a constant flux normalized to the measured 
$R$-band magnitude of the targets.
Comparable  values can  be
expected for the  other XDINSs in our sample  assuming similar optical
spectra and and X-ray to optical flux normalizations.

%_____________
% Luminosities
%
We  have used  the measured  $H$-band flux  upper limits  of  the five
XDINSs  to derive  their corresponding  IR luminosities  upper limits.
Fluxes have been normalized  either for the measured optical parallax
distances \citep[\citeauthor{kaplan2007} \citeyear{kaplan2007}]{mvk2006}  
or for  the  estimated distances  presented in \citet{posselt2006}.  
The  $H$-band
interstellar extinction  correction has been computed  according to the
relation of \citet{fitz1999} using as a reference the $A_V$ inferred
from  the $N_H$  derived  from  the X-ray  spectral  fits in \citet{posselt2006}
and the relations of  Predehl \& Schmitt (1995) and \citet{paresce84}.
In both cases, the computed  $H$-band interstellar extinction is
significantly smaller than the global photometric uncertainty.
% per la 2143 A_H = Paresce: 0.066, P&S: 0.003

We have compared the XDINSs'
IR luminosity upper  limits with the measured IR  luminosities of both
rotation-powered  pulsars  and   magnetars.   For  these  objects,  IR
luminosities have been computed from the measured and the extrapolated
$K$-band magnitudes  \citep[see][for details]{mig2007} using the same 
procedure outlined above.
Although  the comparison
between IR  luminosities obtained in  different bands is  not formally
correct, the $H$ and $K$-bands are close enough in wavelengths so that
the  error due  to the  unknown  color correction  is negligible  with
respect  to  the  overall  luminosity error  budget.   The  comparison
between  different  INS  classes  is   shown  in  Fig.   2,  where  IR
luminosities are plotted vs. the neutron stars' rotational energy loss
$\dot E$ (assuming $I=10^{45}$ g cm$^2$).
\begin{figure}     
\centering    
\includegraphics[bb=10 175 440 600,width=8.0cm,angle=0,clip=]{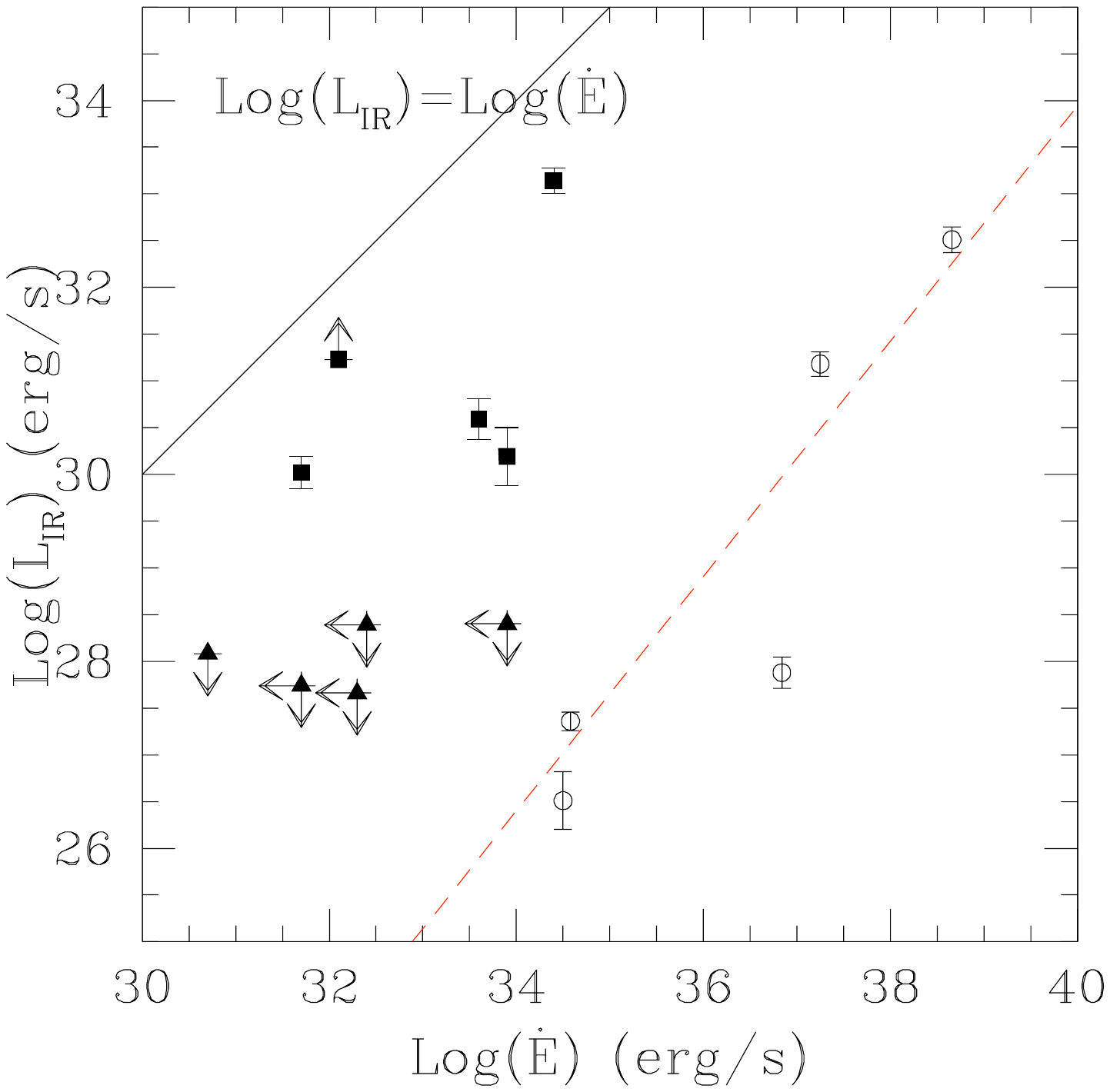}    
\caption{IR luminosities of isolated NSs and derived XDINSs
upper  limits plotted  as a  function  of  the  NSs
rotational energy  loss $\dot E$  \citep[updated from]{mig2007}.
Empty  circles,  filled  squares   and  filled  triangles  indicate
rotation-powered pulsars, magnetars, and XDINSs respectively.  The 
dashed line corresponds to the linear fit for rotation-powered pulsars
while the  solid line shows the  limit case $Log (L_{IR})  = Log (\dot
E)$.   For   the  rotation-powered  pulsars  and   the  magnetars  the
luminosity has been computed in the $K$-band, while for the XDINSs the
luminosity  upper  limits  refer  to  the  $H$-band.   }   \label{Lir}
\end{figure}
As pointed out  by \citet{mig2007},  the correlation between the
IR luminosity  of rotation-powered pulsars and $\dot  E$ suggests that
it is of magnetospheric origin, while the  intrinsically higher IR
luminosity of the  magnetars is likely powered by  the emission from a
fallback  disk and/or by  the magnetic  field decay.   Since
only upper limits on $L_{IR}$ and $\dot E$ (with the only exception of
RX  J0720--3125) are  available for  the XDINSs,  it is  impossible to
recognize a  pattern in the  diagram.  XDINSs could either  follow the
same $L_{IR}$ vs $\dot  E$ correlation of rotation-powered pulsars (or
alike), and  thus be magnetospheric emitter  in the IR,  or they might
not follow any  correlation at all, and thus  be, e.g.  mostly thermal
emitters,  as they  are in  the optical and X-rays.   
Alternatively, they could  represent  an intermediate class
%  of intermediate  IR-bright  neutron  stars
between the magnetars and  the old rotation-powered pulsars, where
all emission processes may coexist. In any case, their low IR emission 
clearly segregates them from magnetars.
% in the $L_{IR}$ vs. $\dot E$ plot.
Whether this is due to an intrinsic difference between the two
classes, or  it is  simply due  to the lower  X-ray luminosity  of the
XDINSs which would reduce the contribution of the putative disk IR emission,
or to a low mass inflow rate through such a disk (or both) is presently unclear.

Searches for fallback disks in different families of neutron stars 
were recently undertaken by several authors, e.g. \citet{wang2007}, \citet{mig2007}
on four compact central objects and on one high magnetic field radio pulsar respectively,
both reporting non-detections. 
Instead \citet{wang2006} successfully detected a flattening in the spectrum 
%(starting in the I band) 
of the AXP 4U 0142+61 from Keck ($K$-band) and Spitzer observations, interpreting 
the data as a detection of a fallback disk of mass $\approx 10^{-5} M_{\odot}$ 
around the neutron star. 
The colours of the detected source are unusually red with respect to main 
sequence stars or giant stars.
Fallback disks could be found in principle around XDINSs as well
\citep[\citeauthor{michel1981} \citeyear{michel1981}]{perna2000}.

To set constraints on the presence of underlying fallback
disks around the XDINSs in our sample
we have used the  derived $H$-band flux upper limits.
Since these objects do not
display radio emission, an hypothetical disk could extend well
inside the light cylinder radius. 
 We note that the detection of pulsed radio emission  at very low frequencies 
(42-112 MHz) has been recently reported for RX J1308.6+2127 and RX J2143.0+0654 \citep{malofeev2007} but has not been independently confirmed so far.
The inner radius of the disk
is generally expected at the magnetospheric radius. Since this
is dependent on the magnetic field and on the accretion rate of the disk 
which is not known a priori, this radius cannot be computed \citep{chn2000}. 
However, given the very low X-ray luminosity level
of the XDINSs, we do not expect that an hypothetical fallback disk
would be accreting on the neutron star, which implies that the minimum value that the
magnetospheric radius can take is the corotation radius. 
Since the corotation radius only depends on the neutron star period, 
it can be computed for all of our objects. 
The disk emission was  computed using the models of \citet{perna2000}, which
include both  the  contribution to  the  emission from  viscous
dissipation,  as well  as the  contribution from  reprocessing  of the
X-ray flux.   However, given  the very low  X-ray luminosity  level of
most XDINSs (see Table 4), we found that, for fluxes below (but close)
to the IR  limits, the contribution to the  disk emission is generally
dominated   by   the  viscous   dissipation   part.   Therefore,   our
observational  limits translate  into limits  on the  maximum possible
mass inflow rate through the  disk. The exception is represented
by RX  J0720.4--3125, which has  the largest X-ray luminosity  and for
which the flux  from a disk with inner edge  at the co-rotation radius
would be instead  dominated by the reprocessing of  the X-ray emission
from the  neutron star.  The expected disk  IR emission would in this case be
brighter than our derived upper limit, unless the disk inner radius is
much larger than the co-rotation radius or the disk has an inclination 
with respect to the line of sight higher than the assumed $60^{\circ}$. 
\begin{table*}[t]
\centering                          % used for centering table
\begin{tabular}{l r r r c c l}        % centered columns (4 columns)
\hline\hline                 % inserts double horizontal lines
Target & $L_X$ & P & $\dot{P}$ & $\dot{M}$ ($R_{in}=R_{cor}$)  & $\dot{M}$ ($R_{in}=R_{lc}$) & Ref. \\    % table heading
       & $10^{31}$ erg s$^{-1}$ & $s$ & $10^{-13} ss^{-1}$ & $10^{-10}M_{\odot}$ year$^{-1}$ & $10^{-10}M_{\odot}$ year$^{-1}$&  \\
\hline                        % inserts single horizontal line
    RX J0420.0--5022 & $3.2$ &  3.45 & $<92$   & 0.56 & 2.06 & 1 \\
    RX J0720.4--3125 & $33.7$ &  8.39 & $0.698$ & - & 3.17 & 2 \\
    RX J0806.4--4123 & $3.5$ & 11.37 & $<18$   & 0.48 & 3.49 & 1 \\
    RX J1856.5--3754 & $ 2.8$ &  7.06 & $<19$   & 0.40 & 2.70 & 3 \\
    RX J2143.0+0654 & $20.3$ &  9.44 & $<60$   & 0.67 & 3.81 & 4 \\

\hline                                   %inserts single line
\end{tabular}
\caption{Estimates of the upper limits of the accretion from a fallback disk. 
The luminosities reported in column two are referred to the distances in column 
three of table \ref{results}.
In column 5 and 6 are shown the upper limits of the accretion rates 
(in units of $10^{-10}M_{\odot}$ year$^{-1}$) for the inner disk radius equal to the 
corotation radius and to the light cylinder radius, respectively. 
The X ray luminosities are obtained from the measured XMM fluxes in the 0.1-2.4 keV energy band \citep{haberl2004b}. 
References: 
(1) \citet{haberl2004};
(2) \citet{zane2002};
(3) \citet{tiengo2007};
(4) \citet{zane2005}
}
\label{mdot}      % is used to refer this table in the text
\end{table*}

The results are shown in Table \ref{mdot}, where the maximum $\dot{M}$ compatible
with the limits is reported. For each object, we derived the limits
using two different values for the inner radius of the disk:
$R_{in}=R_{cor}$ and $R_{in}=R_{lc}$, where $R_{cor}$ is the corotation
radius (the absolute minimum), and $R_{lc}$ is the light cylinder radius 
(for reference). Clearly, 
when the radius is closer in (i.e. when $R_{in}=R_{cor}$), the limits
on $\dot{M}$ are more stringent, due to the fact that there is a larger
contribution of flux coming from the inner regions of the disk.
This translates in an upper limit of the mass inflow though the disk of 
at least $\approx 0.4 \cdot 10^{-10}$M$_{\odot}$year$^{-1}$.

%__________________________________________________________________

\section{Conclusions}

We report on the deepest IR observations ever
of 5 of the 7 known XDINSs. None of the sources was detected in our 
data at the $3 \sigma$ level.
Limiting $H$-band magnitudes are in the $\approx 21.5-22.9$ range.
These limits do not allow to either exclude or confirm a spectral
flattening red-ward of the $R$-band as observed in some magnetars.
%As a consequence, the existance of putative fallback disks, for which
%we derive disk inflow mass rate  upper limits,
%is  also unconfirmed, but not ruled out either.

We  have compared  the derived  IR  luminosity upper  limits with  the
measured  values   of  other   classes  of  isolated   neutron  stars.
Interestingly, we found that 
the distribution of our targets in the $L_{IR}-\dot E$ plane shows
a separation between the XDINSs and the magnetars due to their 
lower (by at least two orders of magnitude) IR luminosity.
Due to the indetermination we have on the IR flux and on $\dot P$ 
we cannot exclude nor confirm that XDINSs will follow a trend similar to the
rotation-powered pulsars in the $L_{IR}-\dot E$ plane,
implying IR emission from the magnetosphere.

We have investigated the possible existence of fallback disks around our 
targets and found that the contribution of the disk emission would be mostly 
dominated by viscous dissipation.
Although we cannot confirm the existence of fallback disks, 
we have derived disk mass inflow rate  upper limits which are
consistent with the age and the disk mass estimates of our targets.

Multi-color deeper IR observations, both in the near infrared and in 
the medium infrared (Spitzer) bands, could better probe the possibility 
of spectral flattening in the IR and its origin.

%\newpage
%__________________________________________________________________

\begin{acknowledgements}
RPM  and GLI acknowledge the  ESO/Chile Scientific Visitors Programme  for
supporting their staying at the ESO Santiago Offices (Vitacura) where  most of this
work  was finalised. RP warmly thanks the ESO/Chile Scientific Visitors Programme for the hospitality during her visit.
GLC thanks the Rome observatory in Monte Porzio Catone for the hospitality.
\end{acknowledgements}

%\begin{thebibliography}{}
%   \bibitem[1997]{zheng} Zheng, W., Davidsen, A. F., Tytler, D. \& Kriss, G. A.
%      1997, preprint
%\end{thebibliography}

\bibliographystyle{aa}
\bibliography{ref}

\end{document}